\documentclass[openacc]{rstransa}

\begin{document}

\title{Emergent gauge symmetries - making symmetry as well as breaking it}

\author{
Steven D. Bass$^{1,2}$
}

\address{$^{1}$Kitzb\"uhel Centre for Physics, Kitzb\"uhel Austria \\
$^2$
Marian Smoluchowski Institute of Physics and 
Institute for Theoretical Physics, Jagiellonian University, Krak\'ow, Poland \\
}

\subject{Particle physics, quantum many-body systems, gauge symmetries, cosmology}


\corres{Steven D. Bass\\
\email{Steven.Bass@cern.ch}}

\begin{abstract}
Gauge symmetries 
play an essential role
in determining the interactions of particle physics. 
Where do they come from?
Might the gauge symmetries of the Standard Model unify in the ultraviolet or might they be emergent in the infrared, 
below some large scale close to the Planck scale?
Emergent gauge symmetries are important in quantum many-body systems in 
quantum phases associated with long range entanglement and topological order,
e.g., 
they arise in high temperature superconductors, 
with string-net condensation and in the A-phase of superfluid $^3$He. 
String-nets and superfluid $^3$He 
exhibit 
emergent properties similar to the building blocks of particle physics.
Emergent gauge symmetries also play an
important role in simulations of quantum field theories.
This article discusses recent thinking on possible emergent gauge symmetries in particle 
physics, commenting also on Higgs phenomena and the vacuum energy or cosmological constant 
puzzle in emergent gauge systems.
\end{abstract}


\begin{fmtext}





\end{fmtext}


\maketitle

\section{Introduction: Key Questions}

Gauge symmetries
underlie modern particle physics determining the interactions of elementary particles
\cite{Pokorski:1987ed,Bjorken:1965zz,Taylor:1976ru}.
Where do gauge symmetries come from?
Do the gauge symmetries of the Standard Model unify in the ultraviolet or might they be 
emergent in the infrared, perhaps below some very large scale close to the Planck scale?
With emergence, the gauge symmetries would "dissolve" above the scale of emergence.

With gravitation, General Relativity can also be viewed as a gauge theory under local 
transformations of the co-ordinate system \cite{Sciama:1964wt,Kibble:1961ba}.
Given the well known challenges with quantising gravity, should it be quantised or might 
General Relativity be emergent at a scale below the Planck scale?

Besides their connection to elementary particle dynamics, the gauge symmetries of particle 
physics are interconnected with Lorentz invariance \cite{Bjorken:1965zz}.
Photons with two physical transverse degrees of freedom are described by a
four component vector $A_{\mu}$.
Whereas the photon (and gluon) fields 
are not 
four-vectors under Lorentz transformations, Lorentz invariance is restored in the action with 
gauge invariance \cite{Bjorken:1965zz,Weinberg:1995mt}.

Gauge invariance is also needed for renormalisability in quantum field theories 
with vector particles \cite{tHooft:1979hnm}.

Local gauge symmetries act on internal degrees of freedom, and not on the physical Hilbert space
which is gauge invariant.
In contrast, global symmetries relate eigenstates of the physical Hamiltonian.

What do we mean by emergence?

Emergence in physics occurs when a many-body system exhibits collective behaviour in the infrared
that is qualitatively different from that of its more primordial constituents as probed in the ultraviolet \cite{Anderson:1972pca}.
Classical physics is emergent from quantum physics.
Protons are emergent from quark-gluon dynamics. Chemistry and biology are emergent from electrodynamics.
As an everyday example of emergent symmetry, consider a carpet which looks flat and translational invariant when looked at from a distance.
Up close, e.g. as perceived by an ant crawling on it, the carpet has structure and this translational invariance is lost.
The symmetry perceived in the infrared, 
e.g. by someone looking at it from a distance, 
``dissolves'' in the ultraviolet when the carpet is observed close up.

Emergent gauge symmetries with spin-one gauge bosons as quasiparticles
are well known in quantum condensed matter physics,
e.g., in quantum phases with long range entanglement and topological order.
Phase transitions come in two types:
the Landau-Ginzburg type with an order parameter corresponding to changes
in some global symmetries and the quantum type, which involve no order parameter
\cite{Spalek:2018aa,Wen:2016ddy}.
Quantum phase transitions are associated with long range entanglement and topological
order
-
see Section 4 below
-
also with restructuring of the Fermi surface.
Examples include the low-energy limit of the 
Fermi-Hubbard model \cite{Baskaran:1987my,Affleck:1988zz}, 
high temperature superconductors, quantum spin liquids and
the fractional quantum Hall effect \cite{Sachdev:2018ddg,Sachdev:2015slk,Powell:2020osu},
string-net condensation \cite{Wen:2004ym,Levin:2004js}
and 
the physics of Fermi points in the A-phase of low temperature superfluid $^3$He 
\cite{Volovik:2003fe,Volovik:2008dd}.
Emergent and "artificial" gauge symmetries are also
employed in quantum simulations of particle physics phenomena, for reviews see 
Refs.~\cite{Banuls:2019bmf,Zohar:2015hwa}.

Emergent Lorentz invariance is also observed in the infrared limit of quantum many-body 
systems starting from a non-relativistic Hamiltonian, though some fine tuning 
\cite{Levin:2005vf}
or 
extra symmetry constraints \cite{Volovik:2003fe}
may be needed to ensure the same effective limiting velocity for all species of quasiparticles.

Emergent gauge symmetries and Lorentz invariance may be connected with infrared fixed points
in the renormalisation group 
-- see Refs.~\cite{Wetterich:2016qee}
and 
\cite{Nielsen:1978is,Chadha:1982qq,Forster:1980dg}.
The universality class and critical dimension are also important.
Might four spacetime dimensions be special for an emergent particle physics Standard Model?

To understand the importance of dimensionality, consider a statistical Ising
system near its critical point.
The long range tail is a renormalisable Euclidean quantum field theory with
properties described by the renormalisation group
\cite{Wilson:1973jj,Jegerlehner:1974dd,Jegerlehner:2013cta,Peskin:1995ev}.
The Landau-Ginzburg criterion tells us that with the Ising model
fluctuations become important for spacetime dimensions of four or less.
This coincides with the dimensionality of space-time.
With four space-time dimensions one finds an interacting quantum field
theory -- $\phi^4$. 
With five or more space-time dimensions the physics 
reduces to a free field theory with long range modes decoupled.
With analyticity the Wick rotation means 
that the Euclidean quantum field theory is 
mathematically equivalent to the theory in Minkowski space.

Might the gauge symmetries of particle physics themselves be emergent?
Here we discuss present thinking on this question with interface to
analogous systems in quantum many-body systems where emergent gauge phenomena are observed to occur.
Might our experience with
quantum many-body systems help understand the possible origin of particle physics symmetries?

We first give an overview of the key issues in particle physics and related phenomenology.

The Standard Model describes our knowledge of particle physics interactions so far measured from low energy precision experiments such as the electron's anomalous magnetic moment $a_e = (g-2)_e/2$
and fine structure constant $\alpha$, up to high energy collisions at the LHC.
The Standard Model is built on 
Quantum Electrodynamics, QED, with
U(1) gauge symmetry,
Quantum Chromodynamics, QCD, with colour SU(3),
and 
chiral SU(2)$_L \otimes$U(1) 
for electroweak interactions with 
the discovery of the Higgs boson ar CERN completing the Standard Model 
\cite{Bass:2021acr}.

The Standard Model is mathematically consistent up to the Planck scale.
Extrapolating with the masses and couplings measured at LHC energies up to the Planck scale,
one finds that the Standard Model
Higgs vacuum sits very close to the border of stable and metastable
(within 1.3 standard deviations of being stable \cite{Bednyakov:2015sca}),
perhaps indicating some critical behaviour in the 
ultraviolet 
\cite{Jegerlehner:2013cta,Degrassi:2012ry,Buttazzo:2013uya}.
The vacuum stability of the Standard Model 
corresponds to whether the Higgs self-coupling $\lambda$ crosses zero 
deep in the ultraviolet and is very sensitive to the values of Standard Model parameters.
The Higgs self-coupling
decreases from its value at LHC energies when evolving to higher scales,
with the evolution dominated by the size of the top quark Yukawa coupling (its coupling to the Higgs boson). 
Without a large top-quark coupling contribution, $\lambda$ would increase as one goes deeper into the ultraviolet and not approach zero.
The Higgs self-coupling
crossing zero, if it happens, occurs at scales bigger than about $10^{10}$ GeV.
The stability of the Higgs vacuum is strongly dependent on the values of Standard Model parameters.
The infrared world of our experiments is then strongly correlated with physics of the ultraviolet.
One finds a possible environmental selection of Standard Model parameters including the 
Higgs mass and couplings with the Standard Model effective field theory supplemented by 
an infrared-ultraviolet correspondence.

While the Standard Model at mass dimension four is mathematically consistent
up to the Planck scale, 
is it also physically valid up to this scale?
One needs some extra physics to understand 
the tiny neutrino masses,
the matter-antimatter asymmetry in the Universe, as well as dark matter, 
the dark energy or cosmological constant and possible primordial inflation.

The key idea of emergence involves the Standard Model as the long range asymptote of a critical system 
that sits close to the Planck scale,
with emergence associated with some
quantum like phase transition deep in the ultraviolet.
Quarks, leptons as well as the
gauge bosons and Higgs boson would then be the stable collective long-range excitations of some 
(unknown) new degrees of freedom that exist above the scale of emergence
\cite{Jegerlehner:2013cta,Jegerlehner:1978nk,Jegerlehner:1998kt,Jegerlehner:2021vqz,Bjorken:2001pe,Bjorken:2010qx,Forster:1980dg}.
With emergence, small gauge groups are most likely preferred.

There are also ideas where the photon may play the role of a Goldstone boson associated with
spontaneously broken Lorentz symmetry, SBLS,
with any effects of SBLS suppressed by 
(at least) 
powers of some large mass scale close to scale of ultraviolet completion,
see Refs.~\cite{Bjorken:1963vg,Bjorken:2010qx}
and
\cite{Chkareuli:2001xe}.
Further ideas consider that fermion quantum field theories 
might be associated at a deeper level with probabilistic cellular automata with classical bits
\cite{Wetterich:2021exk}.

Gauge invariance and renormalisability 
constrain the form of the Standard Model Lagrangian including 
its global symmetries and any global symmetry violations in operator terms up to mass dimension four.
With emergence one finds an additional effective theory like structure with the 
Standard Model Lagrangian supplemented by a tower
of higher dimensional operator terms, 
each suppressed by some power of the large scale of emergence 
\cite{Jegerlehner:2013cta,Bass:2020gpp,Witten:2017hdv}.
In the higher dimensional
operators, which are in general not renormalisable,
new patterns of global symmetry violation are possible.
Especially interesting is lepton number violation 
associated with the dimension five Weinberg operator 
\cite{Weinberg:1979sa}
which may also be the source of tiny neutrino masses 
if neutrinos are Majorana particles, e.g., their own antiparticles.
The Weinberg operator gives neutrino masses
$m_{\rm \nu} 
\sim \Lambda_{\rm ew}^2/M$
where $\Lambda_{\rm ew}$ is the electroweak scale and $M$ is a large ultraviolet scale, here taken as the scale of emergence.
Proton decays might arise at mass dimension six, 
suppressed by two powers of $M$
\cite{Weinberg:1979sa,Wilczek:1979hc}.
Possible axion couplings might enter at mass dimension five in connection with 
the strong CP puzzle in the Standard Model
\cite{Weinberg:1977ma,Wilczek:1977pj},
e.g., the absence of CP violation in QCD 
which could be generated through gluon topological effects and manifest in the neutron electric dipole moment.
Possible axion particles are a candidate for dark matter.
Extra CP violation, perhaps relevant to baryogenesis,
can also appear in high dimensional operators starting at mass dimension six
\cite{Grzadkowski:2010es}.

The cosmological constant measures the vacuum energy perceived by gravitation.
In General Relativity a finite cosmological constant gives no solution to Einstein's equations 
with the Minkowski metric as a global solution
\cite{Weinberg:1988cp}.
Global spacetime invariance is broken by a finite cosmological constant.
If one requires that
the Minkowski metric and global spacetime invariance
as a renormalisation condition at mass dimension four,
then symmetry violations might start at earliest 
with $\Lambda_{\rm ew}/M$ supression, with one suppression 
factor for each dimension of spacetime. 
The scale of the cosmological constant 
induced by Standard Model fields
then comes out to be of the same order as what we expect for light Majorana neutrino masses,
e.g.,
$\mu_{\rm vac} \sim \Lambda_{\rm ew}^2/M$
\cite{Bass:2020egf,Bass:2020nrg}. 
Experimentally, 
the cosmological constant scale
is measured to be 0.002 eV 
from astrophysics experiments \cite{Planck:2018vyg}
corresponding to  $M \sim 10^{16}$ GeV.
The value 0.002 eV is typical of the range expected 
for light neutrino masses \cite{Altarelli:2004cp}.
Next generation neutrino-less double beta decay experiments 
expect to be sensitive to the theoretically interesting mass range 
\cite{Agostini:2017jim,Caldwell:2017mqu}.

With emergence, at
energy scales much bigger than the scale of electroweak symmetry breaking 
Higgs induced mass effects become small perturbations 
so the interactions appear increasingly symmetric. 
Then, as one approaches within a few percent of the scale of emergence 
one becomes sensitive to new higher dimensional operators
which contain extra global symmetry breaking.
The system then becomes 
increasingly chaotic until we go through the emergence phase transition 
to new unknown degrees of freedom above the scale of emergence.
An analogy of this phase transition would be like the transition from quarks to hadrons, 
except here in the deep ultraviolet instead of in the infrared.

Within the emergence scenario, one would like to understand the universality class of the 
Standard Model
and the possible critical dimension
associated with any (quantum) phase transition that produces it.
Might four spacetime dimensions be special 
for an emergent Standard Model?
What is the scale of emergence and might any phase transitions
related to emergent particle physics also be connected with the onset of inflation?
At a deeper level might
gravitation and General Relativity also be emergent?
For the universality class of the Standard Model there are ideas based on 
string-net condensation \cite{Levin:2004js}
and
Fermi points \cite{Volovik:2003fe}
motivated by phenomena in quantum many-body physics.

The plan of this article is as follows.
In Section 2 we give a brief review of gauge symmetries in particle physics.
Section 3 discusses 
ideas about possible emergent gauge symmetry in particle physics.
In Section 4 we discuss
quantum many-body systems
where emergent gauge symmetries are observed
with analogies to phenomena in particle physics.
Lastly,
Section 5 brings together the key ideas and the 
challenges 
with looking for possible emergent phenomena in particle physics.

\section{Gauge symmetries in particle physics and Lorentz invariance}

In this Section we summarise key issues with gauge theories, working to mass dimension four, 
which should also feature in any consistent theory of emergence.

\subsection{Quantum Electrodynamics}

Quantum Electrodynamics, QED, is described through the Lagrangian
\begin{equation}
    {\cal L} = {\bar \psi} \bigl(i 
    \gamma^{\mu} D_{\mu}
    - m\bigr) \psi - \frac{1}{4} F_{\mu \nu} F^{\mu \nu} .
\label{eq:2a}
\end{equation}
Here $\psi$ is the electron field and $A_{\mu}$ denotes the photon.
The gauge covariant derivative 
$
    D_{\mu} \psi = (\partial_{\mu} + ie A_{\mu}) \psi
$
gives
the electron kinetic energy and the electron-photon interaction
with $e$ the electric charge;
$
    F_{\mu \nu} = \partial_{\mu} A_{\nu} - \partial_{\nu} A_{\mu}
$
is the photon field tensor.

The QED Lagrangian can be derived by requiring
invariance under the local U(1) gauge transformation
$
\psi 
\to e^{i \omega(x)} \psi
$
with the kinetic term $\partial_{\mu} \psi$
replaced by the gauge covariant derivative $D_{\mu} \psi$ with the
photon field transforming as
$
A_{\mu} \to A_{\mu} + \frac{1}{e} \partial_{\mu} \omega
$
so that
$ D_{\mu} \psi \to e^{i \omega(x)} D_{\mu} \psi$.
Maxwell's equations are derived from the photon's equations of motion,
\begin{equation}
    \partial_{\mu} 
    F^{\mu \nu} = j^{\nu}
\label{eq:2b}
\end{equation}
with
$j^{\nu} = i e
{\bar \psi} \gamma^{\nu} \psi$,
including Gauss's Law 
${\rm \nabla.E} = \rho$ where 
${\rm E}$ is the electric field
${\bf E} = -
\partial {\bf A}/
{\partial t} - {\bf \nabla} A_0
$
and
$\rho= ie \psi^\dagger \psi$.

Global symmetries are determined by the requirements of gauge invariance
and renormalisability, 
e.g., conserved electric charge corresponding to invariance under global U(1) transformations 
and chiral symmetry with massless electrons.

Gauge invariance is interconnected with Lorentz symmetry.
In general,
$A_{\mu}$ does not transform as a four-vector
under Lorentz transformations
but is supplemented by an additional gauge term.
Let $U(\epsilon)$ denote an infinitesimal unitary Lorentz transformation.
One finds
\begin{equation}
    U(\epsilon) A_{\mu}(x) U^{-1}(\epsilon)
    =
    A_{\mu}(x') - \epsilon_{\mu \nu} A^{\nu} (x')
    +
    \frac{\partial \Lambda(x',\epsilon)}
    {\partial x'^{\mu}} 
\label{eq:2c}
\end{equation}
where
$\Lambda$ is an operator gauge function.
Gauge invariance ensures that the action remains Lorentz invariant.
The structure of Eq.~(\ref{eq:2c}) ensures that
gauge invariant 
Maxwell equations are Lorentz covariant~\cite{Bjorken:1965zz,Weinberg:1995mt}.

Real photons come with two transverse polarisations whereas
$A_{\mu}$ 
has also
time and longitudinal components 
which need not be considered dynamical degrees of freedom.
For example,
canonical quantisation 
singles out $A_0$ at the expense of manifest covariance. 
The time component $A_0$ commutes with all operators and is a c-number and not an operator in contrast with the space components $A_i$.
Gauss's law then implies that
${\rm \nabla. A}$
is also a c-number.
Some selection of "gauge fixing"
(or constraint on the gauge  fields)
defines
the dynamical degrees of freedom in the book keeping of the calculation.
By a suitable choice of gauge - the radiation or Coulomb gauge, 
${\rm \nabla . A} =0$ and $A_0=0$ -
only transverse degrees of
freedom of the photon field
appear in calculations,
though at the expense of manifest Lorentz and gauge covariance in the formalism.
Observables such as S-matrix elements are Lorentz covariant and independent
of gauge.

\subsection{Non-abelian Quantum Chromodynamics}

In QCD Yang-Mills theory the fermions form an SU(3) triplet with the theory invariant
under rotations in SU(3) colour space $\psi \to U \psi$,
where $U$ is an element of the gauge group.
The QCD gauge covariant derivative is
$
    D_{\mu} \psi = \bigl(\partial_{\mu} 
    + i g_3 \frac{\lambda_a}{2}
    A^a_{\mu} \bigr) \psi
$
where $A_{\mu}^a$ are the gluon fields and 
$\lambda_a$ 
are
the SU(3)
Gell-Mann matrices that form the generators of SU(3).
Under SU(3)
gauge transformations
the gauge covariant derivative transforms as
$
D_{\mu} \to U D_{\mu} U^{-1}
$
with 
$
A_{\mu} (x) \rightarrow A_{\mu} \ ' (x) 
=
U A_{\mu} U^{-1}  + \frac{i}{g_3} (\partial_{\mu} U) U^{-1} 
$.
The gluon field tensor 
$    
G_{\mu \nu} = [D_{\mu}, D_{\nu}]_-
$
then induces non-abelian 3 and 4 gluon 
interaction
vertices with the gluons carrying colour charge as well as the quarks.

The 3 gluon interaction vertex, in turn, induces asymptotic freedom whereby
the QCD coupling
$\alpha_s(Q^2) = g_3^2/4 \pi$
decreases logarithmically with increasing resolution
or four-momentum transfer squared 
$Q^2$
with which we probe the QCD system,
in contrast to QED where the renormalised $\alpha$ 
rises slowly with increasing $Q^2$.
With $\alpha_s$ rising in the infrared one finds strongly interacting QCD 
with quark and gluon confinement. Only hadrons,
colourless bound states of quarks and gluons, exist in the ground state spectrum.

QCD Yang Mills also comes with non-trivial gluon topology.
Gauge transformations come in two kinds:
small gauge transformations
(e.g., associated with perturbative QCD)
which are topologically deformable to the identity and large transformations which change the 
topological winding number \cite{Shifman:1988zk}.
The QCD vacuum is a superposition of vacuum states characterised by different topological 
winding number, quantised in integer or fractional units, and delocalised quark chirality 
so that each contributing vacuum state has zero net axial charge.

Under path integral quantisation,
extra Fadeev-Popov ghost fields are introduced to
preserve consistency of the theory and to maintain unitarity.
These ghost fields violate usual spin statistics and 
arise only in loop diagrams as virtual particles with no free states.
The ghost fields are non-physical and 
enter loop diagrams as an artifact of using
the path integral quantisation procedure.

Gauss Law in QCD is important in Hamiltonian lattice QCD
\cite{Creutz:1984mg}.
The net colour electric flux from each lattice site vanishes and ensures that the resulting 
Hilbert space is invariant under small gauge transformations.
For discussion of large gauge transformations, see 
Ref.~\cite{Grosse:1996hp}.

\subsection{Higgs phenomena}

Weak interactions are described by chiral SU(2)$_L$
interactions between left-handed quark and lepton doublets 
mediated by massive W and Z gauge bosons.
Mass terms 
for gauge bosons violate gauge invariance without extra ingredients in the theory.
This problem is resolved through the Brout-Englert-Higgs (BEH) mechanism
\cite{Higgs:1964ia,Higgs:1964pj,Higgs:1966ev,Englert:1964et,Veltman:1997nm}.
The gauge symmetry of the underlying theory can be hidden in the ground state.
The SU(2) gauge bosons couple to an extra complex scalar doublet field $\phi$
which comes with the potential
$
V(\phi) 
= \frac{1}{2} \mu^2 (\phi^* \phi) 
+ \frac{1}{4}
\lambda 
(\phi^* \phi)^2
$.
Here $\lambda >0$ for vacuum stability, 
viz., a minimum of the potential.
The case $\mu^2 < 0$
comes with classical degenerate minima
signaling spontaneous symmetry breaking phenomena.

In the Standard Model this is most transparent when formulated in 
unitary gauge 
where massless Goldstone modes decouple, being  "eaten" to become the longitudinal modes of 
the massive W and Z bosons,
conserving the number of degrees of freedom.
The fourth component of the BEH field $\phi$ is the scalar Higgs particle,
discovered at CERN in 2012 with mass 125 GeV.
The BEH mechanism ensures 
renormalisability \cite{tHooft:1971qjg,tHooft:1972tcz,Veltman:1968ki},
with perturbative unitarity 
ensured with the Higgs mass measured at the LHC.
Conversely, 
consistent high energy behaviour
requires Yang-Mills structure 
with massive gauge bosons
when one goes beyond massive QED 
\cite{LlewellynSmith:1973yud,Bell:1973ex,Cornwall:1973tb,Cornwall:1974km}.

The BEH mechanism comes with a subtlety.
Elitzur's theorem tells us  that there is no gauge invariant order parameter 
associated with the BEH mechanism
(since gauge symmetry does not act on physical states) 
\cite{Elitzur:1975im}.
Spontaneous symmetry breaking is defined 
relative to the choice of gauge, e.g., the unitary gauge,
with all gauge choices being physically equivalent
\cite{Kibble:2014gug}.

There is a related issue
with the definition of asymptotic states.
Gauge symmetries are properties of internal degrees of freedom
with the physical Hilbert space gauge invariant.
This means that
gauge symmetries do not connect physical particle states.
For the Standard Model,
particle states defined via gauge singlet
composite operators involving the 
fermion doublets and gauge bosons 
plus the Higgs doublet were considered in Ref.~\cite{tHooft:1979yoe}
in the context of a "confinement phase" model
and 
Refs.~\cite{Frohlich:1980gj,Frohlich:1981yi}, 
where
the propagators 
corresponding to the gauge singlet composite operators
were found 
to be coincident 
with perturbative Standard Model ones 
at leading order in the Higgs vacuum expectation value. 
Compositeness here 
has nothing to do with real bound states and binding energy. 
The decoration Higgs-ghost fields just produce singlet fields while the Higgs-ghosts 
are transporting away the gauge degrees of freedom. 
This is just the situation we have when working in the unitary gauge. 
These fields are the gauge orbits of the unitary fields and do not change the physics 
at all as we know it. 
The bookkeeping within a calculation looks very different though 
\cite{Jegerlehner:1985ch,Jegerlehner:1984ia}.

Similar issues will arise with Elitzur's theorem and the definition of
quasiparticle states with coupling to emergent Yang-Mills fields 
in quantum many-body systems.

\subsection{The Weinberg-Witten theorem and emergent gauge bosons}

The Weinberg-Witten theorem \cite{Weinberg:1980kq}
is sometimes quoted as a strong constraint
on ideas about possible emergent or composite (massless) gauge bosons.
However, it does come with loopholes 
with exemptions for photons in QED, gluons in QCD, 
W and Z bosons and possible gravitons \cite{Loebbert:2008zz}.

The theorem comes in two parts and states:

1.
A theory that allows the construction of a Lorentz-covariant conserved
four-vector current $J^{\mu}$
cannot contain massless fields
of spin $j> \frac{1}{2}$ with nonvanishing values of the conserved charge $\int J^0 d^3 x$.

2.
A theory that allows the construction of a conserved Lorentz-covariant 
energy-momentum tensor 
$\theta^{\mu \nu}$
for which
$\int \theta^{0 \nu} d^3x$
is the energy-momentum four-vector 
cannot contain massless particles of spin $j >1$.

This theorem is evaded by the gauge bosons of Standard Model interactions.
For QED the photon does not carry the charge of the electromagnetic current.
Weak interactions are mediated by massive W and Z bosons
with masses coming from the BEH mechanism.
The gauge boson masses
get them around the theorem.
In QCD the current either contains just quark degrees of freedom, so that the gluon is not charged under it, or is gauge dependent and therefore not covariant.
There is also the issue of confinement that free massless gluons do not exist in Nature.
With possible gravitons associated with any quantised version of General Relativity,
likewise the graviton not charged under a Lorentz covariant
current.

\section{Emergent Gauge Symmetries and Particle Physics}

Where might our gauge symmetries come from?

With particle physics, 
emergence has the leptons, quarks, gauge and Higgs bosons as
the 
stable long-range
collective excitations of 
some critical statistical system that sits
close to the Planck scale.
Any phase transition associated with emergence 
might be connected
with criticality and the stability of the Standard Model Higgs vacuum.
Physics above the scale of emergence would most likely be described by different degrees of 
freedom with different physical laws.

Motivated by experience in quantum many-body physics, 
one thinks of quantum-like phase transitions
to give 
emergent gauge fields.
If the long range asymptote of a critical Planck system is a renormalisable quantum field 
theory with massless vector particles, then 
renormalisability implies gauge invariance with the massless vector excitations.

As emphasised by Jegerlehner
\cite{Jegerlehner:2013cta,Jegerlehner:2021vqz},
vacuum stability below the phase transition of emergence would
impose constraints on particle masses, including the Higgs mass
so the hierarchy puzzle 
(why the Higgs mass  squared is very much less than the Planck mass squared despite a quadratically divergent counterterm with renormalisation)
may be resolved via environmental selection.
An emergent Standard Model comes with a 
tower of higher dimensional operators with new global symmetry violations suppressed by powers of the scale of emergence.
There is a usual assumption of exact gauge invariance in each term in the higher dimensional operator tower.
Possible gauge symmetry violations in higher dimensional operators are discussed in 
Ref.~\cite{Bjorken:2001pe}.

If the coefficient of the Higgs mass counterterm were to cross zero
with running Standard Model couplings
below Planck scale and below the
scale of emergence,
this would trigger an additional
first order phase transition
which might be important for inflation 
\cite{Jegerlehner:2014mua}.
The scale where this crossing takes place 
is calculation dependent, not below the Planck mass in \cite{Bass:2020nrg}, about 
$10^{16}$ GeV in \cite{Jegerlehner:2013cta}
with a stable vacuum, and
$10^{20}$ GeV in 
\cite{Masina:2013wja} 
and much above the Planck scale in \cite{Degrassi:2012ry}
and
\cite{Hamada:2012bp} with a metastable vacuum.

In the context of possible emergence from a statistical system 
deep in the ultraviolet, 
it is interesting 
also to note that
the Ising model with no external magnetic field has the same 
vacuum equation of state as the cosmological constant 
\cite{Bass:2014lja}.

One can also find emergent symmetries through renormalisation
flow and the decoupling of
heavy ultraviolet
gauge and Lorentz
violating terms in the infrared
(but keeping 
 with same fundamental boson and fermion fields as degrees of freedom)
so that gauge and Lorentz symmetries emerge in connection with an infared fixed point in the renormalisation group.
Here, examples with detailed calculations are given 
by
Wetterich \cite{Wetterich:2016qee}
and by
Nielsen and collaborators \cite{Nielsen:1978is,Chadha:1982qq,Forster:1980dg}.

In other ideas,
Bjorken \cite{Bjorken:1963vg,Bjorken:2001pe}
has considered the  photon as a Nambu-Goldstone boson associated with spontaneous breaking of Lorentz symmetry, SBLS.
Experimentally, Lorentz invariance is known to work to very high accuracy with no deviations observed in experiments
including from low-energy precision measurements (including with  anti-matter) through to the highest energy cosmic ray and neutrino events -- see e.g. 
\cite{Shore:2004sh}.
In Ref.~\cite{Bjorken:1963vg,Bjorken:2001pe}
SBLS is taken
with scale $\approx M$, 
close to the scale of UV completion.
The net vector field
$A_{\mu}$ is taken as 
$A_{\mu} = a_{\mu} + M n_{\mu}$
with 
$a_{\mu}$ is the physical photon field and $n_{\mu}$ is a unit timelike vector
characterising the SBLS.
Motivated by experimental constraints on Lorentz invariance,
possible violations of Lorentz invariance were phenomenologically conjectured 
to be of order
$\delta ({\rm LV}) \sim \mu_{\rm vac}/M
$
where
$\mu_{\rm vac}$
is the cosmological constant scale with
$
\mu_{\rm vac}
 \sim \Lambda_{\rm ew}^2/M$
 with
 $\Lambda_{\rm ew}$ the scale of electroweak symmetry breaking.
\cite{Bjorken:2001pe,Bjorken:2001yv}.
That is, Lorentz violations are here conjectured to arise just as some power of $\mu_{\rm vac}/M$.
With SBLS
the preferred reference frame in this formalism is naturally identified with the frame where the cosmic microwave background is locally at rest.
Note that the cosmological constant scale here 
$
\mu_{\rm vac}
 \sim \Lambda_{\rm ew}^2/M$
is the same as that found with the emergence arguments in Section 1,
with details discussed 
by Bass and Krzysiak in Refs.~\cite{Bass:2020egf,Bass:2020nrg}.

In an alternative approach,
Chkareuli, Froggart and Nielsen
\cite{Chkareuli:2001xe}
considered possible SBLS together with an additional requirement of
non-observability of all Lorentz non-invariant SBLS terms.
Here one again considers
the form
$A_{\mu} = a_{\mu} + M n_{\mu}$
in 
the general Lagrangian density restricted to terms of mass dimension four or less.
The contribution 
$M n_{\mu}$ represents
a classical background field appearing when the vector field $A_{\mu}$ develops a vacuum expectation value.
For QED this involves the extra terms
\begin{equation}
\frac{1}{2} M_A^2 A_{\mu} A^{\mu}
    + \frac{f}{4} A_{\mu} A^{\mu} .
    A_{\nu} A^{\nu}
\label{eq:3a}
\end{equation}
where the mass term $M_A$ and interaction coupling $f$ are to be fixed.
Requiring non-observability of terms involving $n_{\mu}$,
so 
that the net term involving $n$ vanishes,
corresponds to requiring $M^2=f=0$
for the coefficients of the possible gauge symmetry violating terms proportional 
to the square of the vector field.
Otherwise one has to impose an additional gauge fixing term
$n.a=0$ which is inconsistent with the Lorentz gauge 
$\partial^{\mu} a_{\mu}=0$ 
taken as already imposed on the vector field $a_{\mu}$.
This argument can be readily generalised to non-abelian 
gauge fields.
The vector fields  become the source for the gauge symmetry
with the SBLS
converted into gauge degrees of freedom of the 
massless vector fields.
This scenario compares with
the usual picture with gauge invariance being the source for the vector gauge fields.

Besides ideas on the possible emergence of 
Standard Model gauge symmetries below a scale
deep in the ultraviolet,
about 10$^{16}$ GeV or more,
ideas have also been
considered where gauge bosons and perhaps also the
quarks and leptons are constructed as composites of 
some more primordial fermions 
on a distance scale large compared to 
10$^{16}$ GeV -
see e.g. 
Refs.~\cite{Terazawa:1976xx,Fritzsch:1981zh,Fritzsch:2012fv}.
These models tend to predict new states
in the LHC energy range that so far have not been seen in the experiments.

\section{Emergent Gauge Symmetries in Quantum Condensed Matter Systems}

Emergent gauge symmetries are a characteristic of quantum condensed matter systems 
with long range entanglement and topological order.
Here, 
emergent gauge symmetries were first found in the low energy limit of the Fermi-Hubbard model of electron systems by 
Anderson and collaborators
\cite{Baskaran:1987my,Affleck:1988zz}.
This model is important in
discussions of high temperature superconductors
and quantum spin liquids
\cite{Sachdev:2015slk,Sachdev:2018ddg,Powell:2020osu}
as well as in quantum simulations of gauge theories
\cite{Banerjee:2012pg}.

The Fermi-Hubbard model with a two  dimensional lattice serves 
as an important and useful prototype model to explain the key concepts and ideas.
In this Section we first
review how emergent gauge symmetries arise in this system.
With particle physics in mind,
we then
briefly discuss 3+1 dimensional string-nets 
(involving qubit chains in a lattice environment)
explored by
Levin and Wen
\cite{Levin:2004mi,Levin:2004js,Wen:2001yp}
as a model of electrons and photons
with extension to quarks and gluons.
Here one finds emergent gauge fields plus fermions as fluctuations 
and defects of 
long range entanglement.
Finally, 
we describe work by Volovik \cite{Volovik:2003fe,Volovik:2008dd}
on
the A-phase of superfluid $^3$He
where the quasiparticles
include gapless chiral fermions and SU(2) gauge fields
and which may be used for simulating 
many Standard Model particle physics phenomena.
Here one also finds emergent Lorentz invariance and emergent gravity.

Long range entanglement 
\cite{Kitaev:2005dm,Levin:2006zz}
is a universal property of quantum many-body 
systems, insensitive to microscopic details of the Hamiltonian.
The following dancing analogy 
\cite{Wen:2012hm}
is helpful to understand the difference between 
topological order and
symmetry-breaking 
with an
associated order parameter.
First,
with
symmetry-breaking orders, every particle/spin, or
every pair of 
particles/spins, 
dances by itself, and they all
dance in the same way
(corresponding to
the long-range order).
With topological order
one has
a global dance, where every particle/spin is
dancing with every other particle/spin in a collective and 
very organised
fashion, 
not just fermions dancing in pairs.
Global dancing patterns 
are a collective effect 
produced from various local dancing rules
which corresponds to a pattern of long range entanglement. 
Transformations between states with different long range entanglement patterns are described by topological phase transitions.
In the context of quantum computers,
with long range entanglement the time taken for disentanglement  depends on the size of the system, whereas for short range entanglement
it is size independent.
More formal definition in terms of unitary evolution and Hilbert space constructions is given in Refs.~\cite{Chen:2010gda,wen:springer}.

\subsection{The Fermi-Hubbard model and its low energy limit}

The Fermi-Hubbard model describes electron behaviour in an atomic lattice.
One considers a lattice of 
atoms supporting only single a atomic state, which can hold up to two electrons with opposite spins.
The electrons interact with the potential of a static lattice of ions.
One neglects any motion of the ion lattice, being only interested
in interactions of electrons and not in dynamical lattice effects, such as phonons. The electrons interact via Coulomb repulsion.
One assumes that all except the lowest band have very high energies and are, thus, energetically unavailable.
Also, that the remaining band has rotational symmetry.
The electron hopping matrix, 
which describes electron motion from
one lattice site to another,
depends just on the distance between lattice sites.
Finally, one restricts to nearest neighbour interactions
(with underlying matrix elements decreasing fast with increasing distance).

The Fermi-Hubbard model Hamiltonian then has two terms:
a hopping term between nearest neighbour sites 
with coupling strength $t$,
plus an "on-site" Fermi-Hubbard repulsion term $U$,
\begin{equation}
    {\cal H} = - t \sum_{(ij)\sigma} c_{i \sigma}^\dagger c_{j \sigma} + U \sum_i 
    c_{i \uparrow}^\dagger c_{i \uparrow} c_{i \downarrow}^\dagger c_{i \downarrow} .
\label{eq:4a}
\end{equation}
Here a 
square lattice is assumed where $ij$ are nearest neighbour bonds.
$c^\dagger_{i \sigma}$ and $c_{i \sigma}$ 
are the creation and annihilation Fock operators for electrons 
with spin $\frac{1}{2}$
on site $i$.
The first term prefers 
non-localisation whereas the second prefers localisation
with just one electron on each lattice site.
Extra "doping" terms can be included by adding a chemical potential.

In the low-energy Mott limit
$U \gg t$ 
the Fermi-Hubbard system behaves as an insulator.
(With extra doping terms described by adding a chemical potential, it becomes a model for describing high temperature superconductors.)
Treating the hopping term as a perturbation and
keeping the leading term evaluated using Rayleigh-Schr\"odinger perturbation theory, 
the Fermi-Hubbard model
reduces to the Heisenberg magnet Hamiltonian.

For the half filled system one finds
\begin{equation}
    {\cal H}_{\rm eff} =
    J \ \sum_{i,j} 
    (c_{i \alpha} ^{\dagger} \sigma_{\alpha \beta} c_{i \beta})
    .
    (c_{j \alpha} ^{\dagger} \sigma_{\alpha \beta} c_{j \beta})
\label{eq:4b}
\end{equation}
where $J=4t^2/U$,
the 
$\sigma$
denote SU(2) Pauli matrices,
and one has the constraint
$c^{\dagger}_{j \alpha} c_{j  \alpha}=1$.

The strongly correlated electron system can 
be described using
a slave-particle
representation 
using auxiliary fermion and boson operators.
One writes
the $c$-electron Fock operators as
a combination of 
"spinon"
$f$-electrons carrying spin and no electric charge, 
and spinless
"holons"
which carry the electric charge,
that is, with spin-charge separation;
for a review see 
Ref.~\cite{fresard}.
In the low-energy Fermi-Hubbard model
with half filling
the spin operators appearing in the product in Eq,~(\ref{eq:4b})
are chargeless and
the low energy system can be written just in terms of 
the $f$-electrons,
$c \mapsto f$ in Eq.~(\ref{eq:4b}).

The Hamiltonian 
(\ref{eq:4b})
has the important local gauge symmetry 
$f_{j \sigma}^{\dagger}
\to 
e^{i \theta_j}
f_{j \sigma}^{\dagger}$.
The Fermi-Hubbard system exhibits entanglement and quantum correlations in its ground state 
in the $U \gg t$ limit 
\cite{Powell:2020osu}.

The Hamiltonian 
Eq.~(\ref{eq:4b}) 
can also be expressed in the form \cite{Baskaran:1987my}
\begin{equation}
    {\cal H}_{\rm eff} =
    J \sum_{ij} 
    b_{ij}^{\dagger} b_{ij}
\label{eq:4f}
\end{equation}
where
$b_{ij}^{\dagger}
=
(1/\sqrt{2})
(f_{i \uparrow}^{\dagger}
f_{j \downarrow}^{\dagger}
-
f_{i \downarrow}^{\dagger}
f_{j \uparrow}^{\dagger}
)
$
are bosonic 
single (two electron) creation and annihilation operators,
which are important 
in the RVB theory of superconductivity 
\cite{Baskaran:ssc,anderson:prl87}.
These two electron excitations might form a Bose Einstein condensate.
Through Elitzur's theorem 
the thermal average 
$\langle b_{ij} \rangle =0$
at all temperatures.

The model system Eq.~(\ref{eq:4b}) also exhibits a local SU(2) gauge symmetry.
To see this, first consider the electron operators
$(f_1, f_2)$
and
$(f_2^\dagger, -f_1^\dagger)$
which transform as SU(2) spin doublets.
These are combined to form the matrix
\begin{equation}
\Psi =
\left(\begin{array}{cc}
f_1 & f_2
\\
f_2^\dagger &
- f_1^\dagger
\end{array}\right) 
\label{eq:4c}
\end{equation}
which transforms under global SU(2),
$
\Psi \to \Psi g$.
One can define a second local SU(2)
by
$\Psi \to h \Psi$.
Here $g$ and $h$ denote 
SU(2) rotations,
viz.
$e^{i \vec{\sigma}.\vec{\omega}/2}$
where 
$\vec{\sigma}$
denotes the SU(2) Pauli matrices and
$\vec{\omega}$ 
is spacetime independent
for $g$ and spacetime dependent for $h$.
Spin operators for global SU(2)
can be written
${\rm S} = \frac{1}{2} \Psi^\dagger \Psi \sigma^{\rm T}$
where $\sigma^{\rm T}$ is the transpose of $\sigma$.
Since
$
\Psi^\dagger \to g^\dagger \Psi^\dagger h^\dagger
$
it follows that the 
spin operators are invariant under local SU(2).
That is, the Heisenberg interaction
is invariant under local SU(2) gauge transformations
with $h$ denoting an element of the gauge group.

The Hamiltonian in Eq.~(\ref{eq:4b})
can be written in terms of the spin operators 
as
\begin{equation}
    {\cal H}_{\rm eff} 
    =
     J/4 \ \sum_{i,j} 
    ({\rm tr} \ \Psi_i^\dagger \Psi_i \sigma^{\rm T})
    .
     ({\rm tr} \ \Psi_j^\dagger \Psi_j \sigma^{\rm T}) .
\label{eq:4d}
\end{equation}
The local gauge symmetry within the Heisenberg model acts
trivially 
on the spin operators but becomes interesting within the 
large $U$ limit of the Fermi-Hubbard model with electron operators
at half filling. 
This is a 
consequence of the redundancy of parametrising
spin operators by
electron operators.
Note that it is the "spinon" $f$-electrons 
that feel the 
emergent 
SU(2) gauge symmetry here
rather than it
being a 
property of the 
charged
$c$-electrons 
of QED
which appear
in the more general
Fermi-Hubbard Hamiltonian Eq.~(\ref{eq:4a}).

The emergent gauge symmetry seen here comes with an energy barrier.
Local SU(2) gauge invariance is valid up to below the Mott-Hubbard energy gap.
For large but finite $U$ there is an approximate gauge symmetry in the sense that 
it is only broken in the sector of the Hilbert space containing high energy states 
with energies of order $U$.

Adding in a new spin one gauge field
gives the net Lagrangian
\begin{equation}
    {\cal L} = \frac{1}{2} \sum_{j} {\rm tr} \
    \Psi_{j}^\dagger \biggl( i \frac{\partial}{\partial t} + B_{j} \biggr) \Psi_{j} - {\cal H}_{\rm eff} 
\label{eq:4e}
\end{equation}
with the gauge field
$
B = \frac{1}{2} \sigma.{\bf B}$
transforming as
$B \to h ( B + i (\partial/\partial t) ) h^{\dagger}$
under the local SU(2) gauge transformations associated with $h$.
The three components of ${\bf B}$ act as 
Lagrange multipliers and guarantee 
the half filled system with 
constraint of
one particle per site
\cite{Affleck:1988zz}.

We refer to Ref.~\cite{Sachdev:2018ddg} 
for detailed discussion of the phase diagram of the Fermi-Hubbard model including confinement and Higgs phases,
as well as application to 
high temperature superconductors.

\subsection{String-net condensation}

In 3+1 dimensions string-net condensation 
has been
proposed as a 
topological phase
model for emergent
electrons and photons, as well as quarks and gluons.
String-net condensed states are
liquids of fluctuating networks of strings.
A 
condensate is formed from extended
structures rather than particles.
Strings correspond to massless spin-one particles and the ends to fermions
\cite{Levin:2004mi,Levin:2004js,Wen:2001yp}.

The idea starts with a
lattice structure with spin zero bosons
-- qubits.
These can form a string structure with occupied sites
building on an empty "vacuum".
Condensation of these strings can occur as separate to condensation of individual bosons.
One introduces an energy penalty for
strings that end or change type in empty space. 
Strings then arrange themselves to minimise the energy into effective extended objects.
These strings can condense 
beyond any condensation of individual boson sites into a Bose Einstein condensate. 
With excitations above the condensate,
closed strings correspond to spin one gauge bosons and
the ends of open strings correspond
to spin $\frac{1}{2}$ fermions
satisfying 
Fermi-Dirac statistics.
That is, light waves become collective excitations of the string-nets, and
electrons correspond to one end of string.
Strings vibrate in two dimensions.
One obtains
two physical degrees of freedom, just as there are two transverse
photon polarisations.
This picture can be generalised to
an emergent QCD with non-abelian SU(3) gauge degrees of freedom 
\cite{Levin:2004js}.

With string-net condensation,
gauge theory becomes a theory for long range entanglement.
One prediction
is that all fermionic excitations must carry some gauge charges.
Fermions are the topological defects of long range entanglement.

Chiral degrees of freedom remain a more open issue here, 
perhaps related to issues of how to include chiral fermions on the lattice.

In condensed matter physics
in 2+1 dimensions 
string-nets 
may be important for understanding 
quantum spin liquids.
The mineral Herbertsmithite with Kagome lattice structure
was
discovered 
to exhibit properties 
of a quantum spin liquid
\cite{Han:2012a}
(whether gapped or gapless)
with 
possible string-net like structure; 
for recent discussion see Ref.~\cite{wen:springer}.

\subsection{Fermi points and emergent gauge symmetries in $^3$He-A}

The low temperature A-phase of superfluid $^3$He exhibits similar structure
to the particle physics Standard Model.
One finds an emergent gauge symmetry. 
The quasiparticles involve SU(2) gauge bosons as well as gapless chiral fermions 
associated with Fermi points in momentum space
\cite{Volovik:2003fe,Volovik:2008dd}.

The key physics involves the topology of the energy spectrum of fermion
quasiparticles in momentum space.
Fermion quasiparticles are gapped except close to
two Fermi points
$E = 
\pm c {\rm \sigma.p}$
corresponding to 
emergent chiral fermions with limiting velocity $c$ and chirality $\pm 1$.
A Fermi point is a topologically stable hedgehog in momentum space.
Close to these Fermi points 
one finds emergence of massless chiral fermions, 
gauge invariance and gauge bosons,
an emergent gravity and Lorentz invariance.
The ordinary spin degree of freedom is perceived by an inner observer as local SU(2),
like weak isospin in particle physics.
This gauged SU(2) is dynamical as it represents some
collective motion of the fermionic vacuum.
The emergent gauge symmetry appears only in the low energy limit.
With fermion quasiparticles related by global discrete 
symmetries and the action for gauge bosons 
obtained by integrating over fermions,
one can obtain the same 
limiting velocity and same Lorentz invariance for each species of quasiparticle 
\cite{Volovik:2003fe}.

The A-phase of $^3$He exhibits extra phenomena of relativistic quantum field theory such 
as the chiral anomaly.
Further, there is an interesting application 
with the "cosmological constant" for the Helium superfluid. 
For a bubble of $^3$He, the effective cosmological constant
vanishes except for a surface term 
which scales as an
inverse power of the bubble radius,
just as the cosmological constant in cosmology scales as the inverse square of the Hubble radius 
with a de Sitter Universe dominated by dark energy.

\section{Conclusions}

One of big surprises of the LHC is that the particle physics
Standard Model works so well.
The Standard Model is mathematically valid up to the Planck scale.
The Higgs vacuum
without coupling
to any undiscovered new particles
sits very close to the border of stable and metastable
with the Higgs self-coupling 
perhaps crossing zero deep in the ultraviolet.
The vacuum stability depends sensitively on the values of particle masses and Standard Model gauge couplings,
suggesting some infrared-ultraviolet correspondence.

These observations have led to renewed thinking that perhaps the Standard Model gauge symmetries and particles might be emergent below an energy scale close to the Planck scale.
In this case, the usual Standard Model would be supplemented with a tower of higher dimensional operator contributions, like in effective field theory, 
with extra global symmetry breaking in the higher dimensional operators suppressed by powers of the large scale of emergence.
The light neutrino masses have a simple interpretation in this picture if the neutrinos are Majorana particles, as does the cosmological constant.

Gauge symmetries are interconnected with Lorentz invariance.
Any emergent gauge symmetry might perhaps be accompanied by emergent Lorentz invariance or spontaneously broken Lorentz symmetry, with effects of SBLS suppressed by at least some powers of the large scale of emergence.

In seeking to understand the physics of a possible emergent Standard Model, it may be helpful to look at quantum many-body systems involving quantum phase transitions with
long range entanglement where 
the building blocks of particle physics are observed, 
viz., 
emergent gauge symmetries, 
Lorentz invariance 
and spin statistics,
Higgs phenomena
and chiral fermions. 
Emergent gauge symmetries are also an important part of quantum simulations of quantum field theories.

At a deeper level
in
the fundamental structure of matter,
the origin of
"quantum" is also an open puzzle with quantum
physics itself possibly
emergent
\cite{adler:cup,tHooft:2007nis,Wetterich:2009tr}.
For an emergent Standard Model
the degrees of freedom above the scale of emergence remain an open question, perhaps not directly accessible to experiments.

What might be the universality class and critical dimension for an emergent particle physics Standard Model?
How might the scale of emergence be connected to vacuum stability (with the Higgs self-coupling perhaps crossing zero in the deep ultraviolet)
and
the onset of primordial inflation?
Might General Relativity also be emergent below the Planck scale?
Our quest to answer these fundamental
questions concerning the deep structure of matter and spacetime
can only benefit from 
fruitful interaction between particle physics, cosmology and quantum physics.

\vskip6pt

\enlargethispage{20pt}

\competing{The author declares that he has no competing interests.}

\ack{
I thank F. Jegerlehner and
J. Krzysiak for many stimulating discussions on the physics issues discussed in this paper.}




\begin{thebibliography}{9}

\bibitem{Bjorken:1965zz}
  J.~D.~Bjorken and S.~D.~Drell,
  {\it Relativistic quantum fields}
  (McGraw-Hill, 1965)
  
 \bibitem{Pokorski:1987ed}
  S.~Pokorski.
  {\it Gauge Field Theories},  
2nd edition 
(Cambridge Univ. Press, 2000)

\bibitem{Taylor:1976ru}
  J.~C.~Taylor,
  {\it Gauge Theories of Weak Interactions}
(Cambridge Univ. Press, 1976)

\bibitem{Kibble:1961ba}
T.~W.~B.~Kibble,
Lorentz invariance and the gravitational field,
J. Math. Phys. \textbf{2} (1961) 212

\bibitem{Sciama:1964wt}
D.~W.~Sciama,
The Physical Structure of General Relativity,
Rev. Mod. Phys. \textbf{36} (1964) 463
[erratum: Rev. Mod. Phys. \textbf{36} (1964) 1103]

\bibitem{Weinberg:1995mt}
S.~Weinberg,
{\it The Quantum Theory of Fields. Vol. 1: Foundations}
(Cambridge Univ. Press, 1995)

\bibitem{tHooft:1979hnm}
  G.~'t Hooft,
  Why Do We Need Local Gauge Invariance in Theories With Vector Particles? An Introduction,
  NATO Sci.\ Ser.\ B {\bf 59} (1980) 101.

\bibitem{Anderson:1972pca}
  P.~W.~Anderson,
  More Is Different,
  Science {\bf 177} (1972) 393.

\bibitem{Spalek:2018aa}
J. Honig and J. Spalek,
{\it A Primer to the Theory of Critical Phenomena}
(Elsevier, 2018).

\bibitem{Wen:2016ddy}
X.~G.~Wen,
Zoo of quantum-topological phases of matter,
Rev. Mod. Phys. \textbf{89} (2017)  041004

\bibitem{Baskaran:1987my}
  G.~Baskaran and P.~W.~Anderson,
  Gauge theory of high temperature superconductors and strongly correlated Fermi systems,
  Phys.\ Rev.\ B {\bf 37} (1988) 580.
  
  \bibitem{Affleck:1988zz}
  I.~Affleck, Z.~Zou, T.~Hsu and P.~W.~Anderson,
  SU(2) gauge symmetry of the large-U limit of the Hubbard model,
  Phys.\ Rev.\ B {\bf 38} (1988) 745.
  
  \bibitem{Sachdev:2015slk}
  S.~Sachdev,
  Emergent gauge fields and the high temperature superconductors,
  Phil.\ Trans.\ Roy.\ Soc.\ Lond.\ A {\bf 374} (2016) 20150248.
  
  \bibitem{Sachdev:2018ddg}
  S.~Sachdev,
  Topological order, emergent gauge fields, and Fermi surface reconstruction,
  Rept.\ Prog.\ Phys.\  {\bf 82} (2019)   014001.
  
 
\bibitem{Powell:2020osu}
  B.~J.~Powell,
  Emergent particles and gauge fields in quantum matter,
  Contemp.\ Phys.\  {\bf 61} (2020) 96.
 
  
  \bibitem{Levin:2004js}
  M.~A.~Levin and X.~G.~Wen,
  Colloquium: Photons and electrons as emergent phenomena,
  Rev.\ Mod.\ Phys.\  {\bf 77} (2005) 871
  
  
  \bibitem{Wen:2004ym}
  X.~G.~Wen,
  {\it Quantum field theory of many-body systems: From the origin of sound to an origin of light and electrons},
  Oxford, UK: Univ. Pr. (2004) 505 p
  
  
  \bibitem{Volovik:2003fe}
  G.~E.~Volovik,
  {\it The Universe in a helium droplet},
  Int.\ Ser.\ Monogr.\ Phys.\  {\bf 117} (2006) 1
  %
  (Oxford Univ. Press, 2006).
  
  
 \bibitem{Volovik:2008dd}
G.~E.~Volovik,
Emergent physics: Fermi point scenario,
Phil. Trans. Roy. Soc. Lond. A \textbf{366} (2008) 2935.


 \bibitem{Banuls:2019bmf}
  M.~C.~Banuls {\it et al.},
  Simulating Lattice Gauge Theories within Quantum Technologies,
  Eur.\ Phys.\ J.\ D {\bf 74} (2020) 165.
  
  
\bibitem{Zohar:2015hwa}
E.~Zohar, J.~I.~Cirac and B.~Reznik,
Quantum Simulations of Lattice Gauge Theories using Ultracold Atoms in Optical Lattices,
Rept. Prog. Phys. \textbf{79} (2016) 014401  
  
   
\bibitem{Levin:2005vf}
  M.~Levin and X.~G.~Wen,
  Quantum ether: photons and electrons from a rotor model,
  Phys.\ Rev.\ B {\bf 73} (2006) 035122
 
 
 \bibitem{Wetterich:2016qee}
  C.~Wetterich,
  Gauge symmetry from decoupling,
  Nucl.\ Phys.\ B {\bf 915} (2017) 135.
  
 
\bibitem{Forster:1980dg}
  D.~Forster, H.~B.~Nielsen and M.~Ninomiya,
  Dynamical Stability of Local Gauge Symmetry: Creation of Light from Chaos,
  Phys.\ Lett.\  B {\bf 94} (1980) 135.

\bibitem{Nielsen:1978is}
H.~B.~Nielsen and M.~Ninomiya,
Beta Function in a Noncovariant {Yang-Mills} Theory,
Nucl. Phys. B \textbf{141} (1978) 153.

\bibitem{Chadha:1982qq}
  S.~Chadha and H.~B.~Nielsen,
  Lorentz Invariance As A Low-energy Phenomenon,
  Nucl.\ Phys.\ B {\bf 217} (1983) 125.

 \bibitem{Wilson:1973jj}
  K.~G.~Wilson and J.~B.~Kogut,
  The Renormalization group and the epsilon expansion,
  Phys.\ Rept.\  {\bf 12} (1974) 75.
  
  
\bibitem{Peskin:1995ev}
  M.~E.~Peskin and D.~V.~Schroeder,
  {\it An introduction to quantum field theory}
	(Westview Press, 1995). 

 
\bibitem{Jegerlehner:1974dd}
  F.~Jegerlehner,
  Quantum Field Theory and Statistical Mechanics,
 Lect.\ Notes Phys.\  {\bf 37} (1975) 114.
 
\bibitem{Jegerlehner:2013cta}
  F.~Jegerlehner,
  The Standard model as a low-energy effective theory: what is triggering the Higgs mechanism?,
  Acta Phys.\ Polon.\ B {\bf 45} (2014)   1167.
  
 
\bibitem{Bass:2021acr}
  S.~D.~Bass, A.~De Roeck and M.~Kado,
  The Higgs boson -- its implications and prospects for future discoveries,
  Nature Rev. Phys. \textbf{3} (2021) 608,
arXiv:2104.06821 [hep-ph].


\bibitem{Bednyakov:2015sca}
A.~V.~Bednyakov, B.~A.~Kniehl, A.~F.~Pikelner and O.~L.~Veretin,
Stability of the Electroweak Vacuum: Gauge Independence and Advanced Precision,
Phys. Rev. Lett. \textbf{115} (2015) 201802


\bibitem{Degrassi:2012ry}
G.~Degrassi, S.~Di Vita, J.~Elias-Miro, J.~R.~Espinosa, G.~F.~Giudice, G.~Isidori and A.~Strumia,
Higgs mass and vacuum stability in the Standard Model at NNLO,
JHEP \textbf{08} (2012) 098


\bibitem{Buttazzo:2013uya}
D.~Buttazzo, G.~Degrassi, P.~P.~Giardino, G.~F.~Giudice, F.~Sala, A.~Salvio and A.~Strumia,
Investigating the near-criticality of the Higgs boson,
JHEP \textbf{12} (2013) 089


\bibitem{Jegerlehner:1978nk}
  F.~Jegerlehner,
  The Vector Boson and Graviton Propagators in the Presence of Multipole Forces,
  Helv.\ Phys.\ Acta {\bf 51} (1978) 783. 
  
  \bibitem{Jegerlehner:1998kt}
  F.~Jegerlehner,
  The 'Ether world' and elementary particles,
  hep-th/9803021.


\bibitem{Jegerlehner:2021vqz}
F.~Jegerlehner,
The Standard Model of Particle Physics as a Conspiracy Theory and the Possible Role of the Higgs Boson in the Evolution of the Early Universe,
Acta Phys. Polon. B \textbf{52} (2021) 575.


\bibitem{Bjorken:2001pe}
  J.~Bjorken,
  Emergent gauge bosons,
  hep-th/0111196. 


\bibitem{Bjorken:2010qx}
  J.~D.~Bjorken,
  Emergent Photons and Gravitons: The Problem of Vacuum Structure,
  arXiv:1008.0033 [hep-ph].


\bibitem{Bjorken:1963vg}
  J.~D.~Bjorken,
  A Dynamical origin for the electromagnetic field,
  Annals Phys.\  {\bf 24} (1963) 174.
  
  
\bibitem{Chkareuli:2001xe}
  J.~L.~Chkareuli, C.~D.~Froggatt and H.~B.~Nielsen,
  Lorentz invariance and origin of symmetries,
  Phys.\ Rev.\ Lett.\  {\bf 87} (2001) 091601
  
 
\bibitem{Wetterich:2021exk}
C.~Wetterich,
Quantum fermions from classical bits,
arXiv:2106.15517 [quant-ph], this volume.
 
  
\bibitem{Bass:2020gpp}
  S.~D.~Bass,
  Emergent Gauge Symmetries and Particle Physics,
  Prog.\ Part.\ Nucl.\ Phys.\  {\bf 113} (2020) 103756.


\bibitem{Witten:2017hdv}
  E.~Witten,
  Symmetry and Emergence,
  Nature Phys.\  {\bf 14} (2018) 116


\bibitem{Weinberg:1979sa}
  S.~Weinberg,
  Baryon and Lepton Nonconserving Processes,
  Phys.\ Rev.\ Lett.\  {\bf 43} (1979) 1566.


\bibitem{Wilczek:1979hc}
F.~Wilczek and A.~Zee,
Operator Analysis of Nucleon Decay,
Phys. Rev. Lett. \textbf{43} (1979) 1571


\bibitem{Weinberg:1977ma}
S.~Weinberg,
A New Light Boson?,
Phys. Rev. Lett. \textbf{40} (1978) 223


\bibitem{Wilczek:1977pj}
F.~Wilczek,
Problem of Strong  $P$  and  $T$  Invariance in the Presence of Instantons,
Phys. Rev. Lett. \textbf{40} (1978) 279.


\bibitem{Grzadkowski:2010es}
B.~Grzadkowski, M.~Iskrzynski, M.~Misiak and J.~Rosiek,
Dimension-Six Terms in the Standard Model Lagrangian,
JHEP \textbf{10} (2010) 085


\bibitem{Weinberg:1988cp}
  S.~Weinberg,
 The Cosmological Constant Problem,
  Rev.\ Mod.\ Phys.\  {\bf 61} (1989) 1.
  
  
  \bibitem{Bass:2020egf}
  S.~D.~Bass and J.~Krzysiak,
  Vacuum energy with mass generation and Higgs bosons,
  Phys.\ Lett.\ B {\bf 803} (2020) 135351.


\bibitem{Bass:2020nrg}
  S.~D.~Bass and J.~Krzysiak,
  The cosmological constant and Higgs mass with emergent gauge symmetries,
  Acta Phys.\ Polon.\ B {\bf 51} (2020) 1251.
  

\bibitem{Planck:2018vyg}
N.~Aghanim \textit{et al.} [Planck Collaboration],
Planck 2018 results. VI. Cosmological parameters,
Astron. Astrophys. \textbf{641} (2020), A6
[erratum: Astron. Astrophys. \textbf{652} (2021), C4]


\bibitem{Altarelli:2004cp}
  G.~Altarelli,
  Neutrino 2004: Concluding talk,
  Nucl.\ Phys.\ Proc.\ Suppl.\  {\bf 143} (2005) 470.
  
  
 \bibitem{Agostini:2017jim}
  M.~Agostini, G.~Benato and J.~Detwiler,
  Discovery probability of next-generation neutrinoless double-$\beta$ decay experiments,
  Phys.\ Rev.\ D {\bf 96} (2017) 053001.


\bibitem{Caldwell:2017mqu}
  A.~Caldwell, A.~Merle, O.~Schulz and M.~Totzauer,
  Global Bayesian analysis of neutrino mass data,
  Phys.\ Rev.\ D {\bf 96} (2017) 073001.

  
\bibitem{Shifman:1988zk}
M.~A.~Shifman,
Anomalies and Low-Energy Theorems of Quantum Chromodynamics,
Phys. Rept.
\textbf{209} (1991) 341


\bibitem{Creutz:1984mg}
  M.~Creutz,
  {\it Quarks, gluons and lattices}
  (Cambridge Univ. Press, 1985)


\bibitem{Grosse:1996hp}
H.~Grosse, E.~Langmann and E.~Raschhofer,
The Luttinger-Schwinger model,
Annals Phys. \textbf{253} (1997) 310.


\bibitem{Higgs:1964ia}
  P.~W.~Higgs,
  Broken symmetries, massless particles and gauge fields,
  Phys.\ Lett.\  {\bf 12} (1964) 132.


\bibitem{Higgs:1964pj}
  P.~W.~Higgs,
  Broken Symmetries and the Masses of Gauge Bosons,
  Phys.\ Rev.\ Lett.\  {\bf 13} (1964) 508.


\bibitem{Higgs:1966ev}
  P.~W.~Higgs,
  Spontaneous Symmetry Breakdown without Massless Bosons,
  Phys.\ Rev.\  {\bf 145} (1966) 1156.


\bibitem{Englert:1964et}
  F.~Englert and R.~Brout,
  Broken Symmetry and the Mass of Gauge Vector Mesons,
  Phys.\ Rev.\ Lett.\  {\bf 13} (1964) 321.


\bibitem{Veltman:1997nm}
  M.~J.~G.~Veltman,
  {\it Reflections on the Higgs system},
  CERN-97-05, CERN-YELLOW-97-05.

  
\bibitem{tHooft:1971qjg}
  G.~'t Hooft,
  Renormalizable Lagrangians for Massive Yang-Mills Fields,
  Nucl.\ Phys.\ B {\bf 35} (1971) 167.
  
\bibitem{tHooft:1972tcz}
  G.~'t Hooft and M.~J.~G.~Veltman,
  Regularization and Renormalization of Gauge Fields,
  Nucl.\ Phys.\ B {\bf 44} (1972) 189.
  
\bibitem{Veltman:1968ki}
  M.~J.~G.~Veltman,
  Perturbation theory of massive Yang-Mills fields,
  Nucl.\ Phys.\ B {\bf 7} (1968) 637.
  
  
\bibitem{LlewellynSmith:1973yud}
  C.~H.~Llewellyn Smith,
  High-Energy Behavior and Gauge Symmetry,
  Phys.\ Lett.\  B {\bf 46} (1973) 233.

\bibitem{Bell:1973ex}
  J.~S.~Bell,
  High-energy Behavior Of Tree Diagrams In Gauge Theories,
  Nucl.\ Phys.\ B {\bf 60} (1973) 427.

  
\bibitem{Cornwall:1973tb}
  J.~M.~Cornwall, D.~N.~Levin and G.~Tiktopoulos,
  Uniqueness of spontaneously broken gauge theories,
  Phys.\ Rev.\ Lett.\  {\bf 30} (1973) 1268
   Erratum: [Phys.\ Rev.\ Lett.\  {\bf 31} (1973) 572].
  
\bibitem{Cornwall:1974km}
  J.~M.~Cornwall, D.~N.~Levin and G.~Tiktopoulos,
  Derivation of Gauge Invariance from High-Energy Unitarity Bounds on the S Matrix,
  Phys.\ Rev.\ D {\bf 10} (1974) 1145
   Erratum: [Phys.\ Rev.\ D {\bf 11} (1975) 972].


\bibitem{Elitzur:1975im}
S.~Elitzur,
Impossibility of Spontaneously Breaking Local Symmetries,
Phys. Rev. D \textbf{12} (1975) 3978.


\bibitem{Kibble:2014gug}
  T.~W.~B.~Kibble,
  Spontaneous symmetry breaking in gauge theories,
  Phil.\ Trans.\ Roy.\ Soc.\ Lond.\ A {\bf 373} (2014) 20140033.


\bibitem{tHooft:1979yoe}
  G.~'t Hooft,
  Which Topological Features of a Gauge Theory Can Be Responsible for Permanent Confinement?,
  NATO Sci.\ Ser.\ B {\bf 59} (1980) 117.

 
 \bibitem{Frohlich:1980gj}
  J.~Frohlich, G.~Morchio and F.~Strocchi,
  Higgs Phenomenon Without A Symmetry Breaking Order Parameter,
  Phys.\ Lett.\  B {\bf 97} (1980) 249.

\bibitem{Frohlich:1981yi}
  J.~Frohlich, G.~Morchio and F.~Strocchi,
  Higgs Phenomenon Without Symmetry Breaking Order Parameter,
  Nucl.\ Phys.\ B {\bf 190} (1981) 553.
   
  
\bibitem{Jegerlehner:1985ch}
  F.~Jegerlehner and J.~Fleischer,
  High-energy Behavior Of The Electromagnetic Singlet Current In The Glashow-Weinberg-Salam Model,
  Phys.\ Lett.\  B {\bf 151} (1985) 65.

  
 \bibitem{Jegerlehner:1984ia}
  F.~Jegerlehner and J.~Fleischer,
  Singlet Form-factors and Local Observables in the Glashow-{Weinberg-Salam} Model,
  Acta Phys.\ Polon.\ B {\bf 17} (1986) 709. 
 
\bibitem{Weinberg:1980kq}
S.~Weinberg and E.~Witten,
Limits on Massless Particles,
Phys. Lett. B \textbf{96} (1980) 59

  
\bibitem{Loebbert:2008zz}
  F.~Loebbert,
  The Weinberg-Witten theorem on massless particles: An Essay,
  Annalen Phys.\  {\bf 17} (2008) 803.


\bibitem{Jegerlehner:2014mua}
F.~Jegerlehner,
Higgs inflation and the cosmological constant,
Acta Phys. Polon. B \textbf{45} (2014) 1215


\bibitem{Masina:2013wja}
I.~Masina and M.~Quiros,
On the Veltman Condition, the Hierarchy Problem and High-Scale Supersymmetry,
Phys. Rev. D \textbf{88} (2013) 093003


\bibitem{Hamada:2012bp}
  Y.~Hamada, H.~Kawai and K.~y.~Oda,
  Bare Higgs mass at Planck scale,
  Phys.\ Rev.\ D {\bf 87} (2013) 053009;
   Erratum: [Phys.\ Rev.\ D {\bf 89} (2014) 059901].


\bibitem{Bass:2014lja}
S.~D.~Bass,
The Cosmological Constant Puzzle: Vacuum Energies from QCD to Dark Energy,
Acta Phys. Polon. B \textbf{45} (2014) 1269.


\bibitem{Shore:2004sh}
  G.~M.~Shore,
  Strong equivalence, Lorentz and CPT violation, anti-hydrogen spectroscopy and gamma-ray burst polarimetry,
  Nucl.\ Phys.\ B {\bf 717} (2005) 86
  
  
  \bibitem{Bjorken:2001yv}
  J.~D.~Bjorken,
  Standard model parameters and the cosmological constant,
  Phys.\ Rev.\ D {\bf 64} (2001) 085008
 
 
 \bibitem{Terazawa:1976xx}
H.~Terazawa, K.~Akama and Y.~Chikashige,
Unified Model of the Nambu-Jona-Lasinio Type for All Elementary Particle Forces,
Phys. Rev. D \textbf{15} (1977) 480
  

\bibitem{Fritzsch:1981zh}
H.~Fritzsch and G.~Mandelbaum,
Weak Interactions as Manifestations of the Substructure of Leptons and Quarks,
Phys. Lett. B \textbf{102} (1981) 319
  

\bibitem{Fritzsch:2012fv}
H.~Fritzsch,
The Size of the Weak Bosons,
Phys. Lett. B \textbf{712} (2012) 231.


\bibitem{Banerjee:2012pg}
  D.~Banerjee, M.~Dalmonte, M.~Muller, E.~Rico, P.~Stebler, U.-J.~Wiese and P.~Zoller,
  Atomic Quantum Simulation of Dynamical Gauge Fields coupled to Fermionic Matter: From String Breaking to Evolution after a Quench,
  Phys.\ Rev.\ Lett.\  {\bf 109} (2012) 175302


\bibitem{Wen:2001yp}
X.~G.~Wen,
Origin of gauge bosons from strong quantum correlations,
Phys. Rev. Lett. \textbf{88} (2002) 011602

 
\bibitem{Levin:2004mi}
M.~A.~Levin and X.~G.~Wen,
String net condensation: A Physical mechanism for topological phases,
Phys. Rev. B \textbf{71} (2005) 045110


\bibitem{Kitaev:2005dm}
A.~Kitaev and J.~Preskill,
Topological entanglement entropy,
Phys. Rev. Lett. \textbf{96} (2006) 110404


\bibitem{Levin:2006zz}
M.~Levin and X.~G.~Wen,
Detecting Topological Order in a Ground State Wave Function,
Phys. Rev. Lett. \textbf{96} (2006) 110405


\bibitem{Wen:2012hm}
X.~G.~Wen,
Topological order: from long-range entangled quantum matter to an unification of light and electrons,
ISRN Cond. Matt. Phys. \textbf{2013} (2013), 198710,
arXiv:1210.1281 [cond-mat.str-el].


\bibitem{Chen:2010gda}
X.~Chen, Z.~C.~Gu and X.~G.~Wen,
Local unitary transformation, long-range quantum entanglement, wave function renormalization, and topological order,
Phys. Rev. B \textbf{82} (2010) 155138


\bibitem{wen:springer}
B. Zeng, X. Chen, D.-L. Zhou, X.-G. Wen,
{\it
Quantum Information Meets Quantum Matter -- From Quantum Entanglement to Topological Phase in Many-Body Systems
}
(Springer, 2019).


\bibitem{fresard}
R. Fr\'esard,
The Slave-Boson Approach to
Correlated Fermions,
Chapter 9 of
{\it
Many-Body Physics: From Kondo to Hubbard
Modeling and Simulation, Vol. 5},
eds. E. Pavarini, E. Koch, and P. Coleman
(Verlag des Forschungszentrum J\"ulich, 2015).


\bibitem{Baskaran:ssc}
G. Baskaran, Z. Zou and P. W. Anderson,
The Resonating Valence Bond State and High $T_c$
Superconductivity - A Mean Field Theory,
Solid State Comm. {\bf 63} (1987) 973.


\bibitem{anderson:prl87}
P. W. Anderson, G. Baskaran, Z. Zou, and T. Hsu,
Resonating —Valence-Bond Theory of Phase Transitions and Superconductivity
in La$_2$Cu0$_4$-Based Compounds,
Phys. Rev. Lett. {\bf 58} (1987) 2790.


\bibitem{Han:2012a}
T. H. Han, J. Helton, 
S. Chu et al., Fractionalized excitations in the spin-liquid state of a kagome-lattice antiferromagnet. 
Nature {\bf 492} (2012) 406. 
%

\bibitem{adler:cup}
S. L. Adler,
{\it
Quantum Theory as an Emergent Phenomenon
}
(Cambridge Univ. Press, 2004).

\bibitem{tHooft:2007nis}
G.~'t Hooft,
Emergent Quantum Mechanics and Emergent Symmetries,
AIP Conf. Proc. \textbf{957} (2007) 154

\bibitem{Wetterich:2009tr}
C.~Wetterich,
Quantum mechanics from classical statistics,
Annals Phys. \textbf{325} (2010) 852.

\end{thebibliography}
\end{document}